\documentclass[amsmath,amssymb,aps,prx,reprint]{revtex4-2}

\usepackage{graphicx}
\usepackage{dcolumn}
\usepackage{bm}
\usepackage{amssymb,amstext,amsmath}
\usepackage{xr}
\usepackage{color}
	 \definecolor{darkred}{rgb}{0.75,0,0}
	 \definecolor{darkgreen}{rgb}{0,0.5,0}
	 \definecolor{darkblue}{rgb}{0,0,0.75}
  	 \definecolor{darkorange}{rgb}{1,0.9,0.1}
	 \definecolor{dark}{rgb}{0,0,0}

\begin{document}

\preprint{APS/123-QED}

\title{Stability of complex communities: A perspective from discrete-time dynamics}


\author{Shuaiying Wang$^{1,\dagger}$}
\author{Yuguang Yang$^{1,2,\dagger}$}%
\author{Aming Li$^{1,3,*}$}
\affiliation{%
\rm
$1$ Center for Systems and Control, College of Engineering, Peking University, Beijing 100871, China\\
$2$ Department of Civil and Environmental Engineering, Massachusetts Institute of Technology, 77 Massachusetts Avenue, Cambridge, MA 02139, USA\\
$3$ Center for Multi-Agent Research, Institute for Artificial Intelligence, Peking University, Beijing 100871, China\\
${\dagger}$ These authors contributed equally to this work.\\
$*$ Corresponding author:~amingli@pku.edu.cn
}%




\begin{abstract}
Understanding the stability of complex communities is a central focus in ecology, many important theoretical advancements have been made to identify drivers of ecological stability. However, previous results often rely on the continuous-time dynamics, assuming that species have overlapping generations. In contrast, numerous real-world communities consist of species with non-overlapping generations, whose quantitative behavior can only be precisely represented by discrete-time dynamics rather than continuous ones. Here, we develop a theoretical framework and propose a metric to quantify the stability of complex communities characterized by non-overlapping generations and diverse interaction types. In stark contrast to existing results for overlapping generations, we find that increasing self-regulation strength first stabilizes and then destabilizes complex communities. This pattern is further confirmed in both exploitative (\emph{E. aerogenes}, \emph{P. aurantiaca}, \emph{P. chlororaphis}, \emph{P. citronellolis}) and competitive (\emph{P. putida}, \emph{P. veroni}, \emph{S. marcescens}) soil microbial communities. Moreover, we show that communities with diverse interaction types become the most stable, which is corroborated by empirical mouse microbial networks. Furthermore, we reveal that the prevalence of weak interactions can stabilize communities, which is consistent with findings from existing microbial experiments. Our analyses of complex communities with non-overlapping generations provide a more comprehensive understanding of ecological stability and informs practical strategies for ecological restoration and control.
\end{abstract}

\maketitle

\section{Introduction}
Stability, capturing whether a system can recover from external perturbations, is a key property of complex ecological communities \cite{may2001stability,may2007theoretical}. The loss of ecological stability may lead to massive species extinctions and potential community collapse, resulting in catastrophic consequences for humanity and the natural world. Therefore, studying the stability of complex communities lies at the heart of ecology. Since the pioneering work of May \cite{may1972nature}, ecologists have performed a tremendous amount of theoretical work to understand and clarify the contributions of different community characteristics to the stability of complex communities. These characteristics mainly cover interaction types \cite{allesina2012nature,mougi2012diversity}, interaction strength \cite{allesina2012nature,allesina2015stability,tang2014correlation}, interaction network structure \cite{newman2006modularity,grilli2016modularity}, spatial structure \cite{gravel2016stability,baron2020dispersal}, and time delays in species interactions \cite{may1973time,niebur1991collective,jirsa2004will,pigani2022delay,yang2023time}.

\begin{figure*}
\centering
\includegraphics[width=0.8\textwidth]{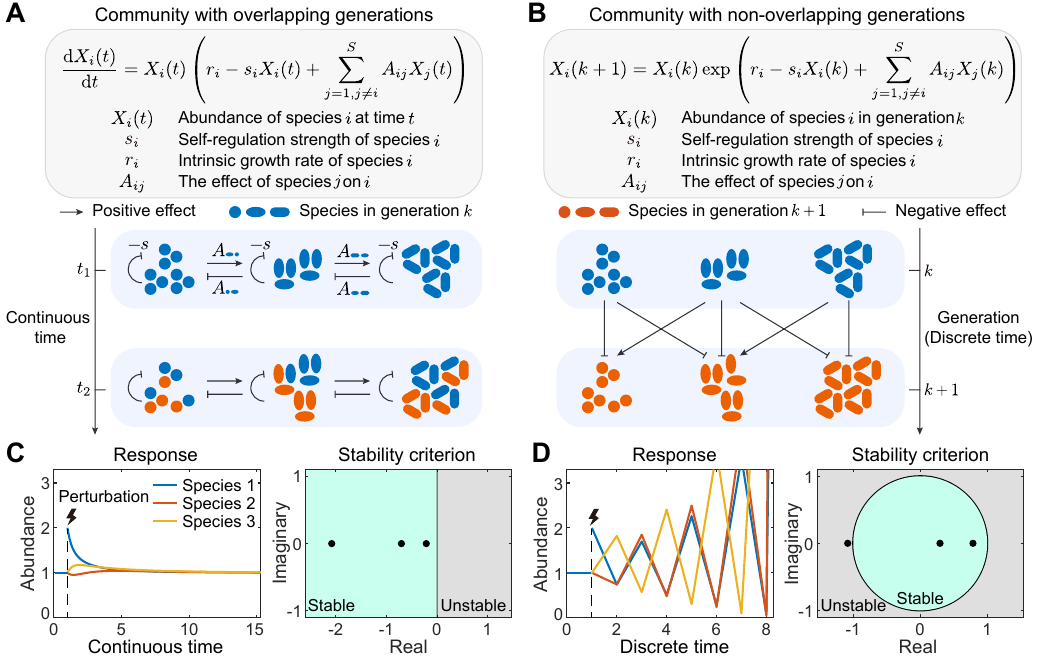}
\caption{
\textbf{Illustration of the stability of complex communities with overlapping and non-overlapping generations.}
\textbf{A}, Schematic diagram of a three--species community with overlapping generations. Blue and red rods represent species from different generations, where sharp-head and blunt-head arrows represent positive and negative effects between species, respectively. For the continuous-time dynamics shown in the grey frame, species from different generations can coexist. \textbf{B}, For the counterpart with non-overlapping generations, species from different generations can not coexist, and the discrete-time dynamics are shown in the grey frame. \textbf{C}-\textbf{D}, Responses of three-species communities to external perturbations (left half of each panel) and eigenvalue distributions of Jacobian matrices (right half of each panel). Communities that recover to their previous state following perturbations are stable, thus community with overlapping generations is stable (C). Mathematically, a community with overlapping generations is stable when all eigenvalues of Jacobian matrix have negative real part, that is, all eigenvalues locate in the left half of the complex plane (green region). In D, when the same community is under discrete-time dynamics, species extinct and this community is no longer stable. Communities under discrete-time dynamics are stable when all the eigenvalues of the Jacobian matrices locate in the unit circle centered at the origin of the complex plane (green region), while here the three eigenvalues are not all in the circle. In C-D, we set $s=1$, $r_1=0.7$, $r_2=0.6$, $r_3=-0.3$, $A_{12}=A_{21}=-1$, $A_{13}=A_{31}=0$, $A_{23}=A_{32}=1$.
}
\label{overlapping_nonoverlapping}
\end{figure*}

However, the vast majority of existing research relies on a crucial assumption that communities are composed of species with overlapping generations. In other words, individuals of the same species from different generations can coexist at the same time, and the dynamical behavior of such communities is described by continuous-time dynamics (Fig.~\ref{overlapping_nonoverlapping}A). While it is true that species with overlapping generations are widespread in nature, there is also a large number of species having non-overlapping generations, where individuals of the same species cannot coexist at the same time (Fig.~\ref{overlapping_nonoverlapping}B). These species, including annual plants \cite{jensen1994dynamics}, univoltine insects \cite{aalberg2012evolution} (i.e., insects only breeding once a year), pink salmon \cite{kovach2013temporal}, are important components of natural ecological communities. For communities composed of such species, continuous-time models cannot accurately capture their behavior, and discrete-time models should be used instead \cite{may1974biological}. As a result, most conventional ecological theory, which was developed for communities of species with overlapping generations, becomes less effective in understanding the stability of such communities.

Theorists have long recognized the short-comings of continuous-time ecological theory and have made great efforts to establish a discrete-time framework and thus to understand the dynamics of communities of species with non-overlapping generations. During the early period, ecologists paid much attention to species-poor communities. In 1974, May's prominent work studied the dynamics of a single population with non-overlapping generations and found that stability loss of such populations can lead to surprising dynamical behaviors like limit cycles and chaos \cite{may1974biological}. Beddington \textit{et al.} extended May's work to a two-species predator-prey model for populations with non-overlapping generations and showed that the loss of stability can also lead to similar patterns \cite{beddington1975dynamic}. Recently, researchers have begun to focus on species-rich communities. Sinha \textit{et al.} focused on communities with randomly assigned interactions and demonstrated that increasing the number of interactions per species or the intensity of interactions leads to an increased likelihood of species extinctions \cite{sinha2005evidence}. However, in addition to species richness, many other properties--such as interaction type composition, interaction strength distribution, and network structure--can emerge in real-world complex communities. For communities composed of species with non-overlapping generations, we still lack a comprehensive understanding of how these properties influence stability.

Inspired by this gap, here we develop a new framework to analyze the stability of large complex communities composed of species under discrete-time dynamics. This allows us to explore in detail the influence of different global community characteristics on stability by integrating theoretical analyses with empirical data. Surprisingly, we find that several key principles governing ecological stability need to be modified. These findings highlight the importance of studying communities of species with non-overlapping generations to comprehensively understand and control complex communities.

\begin{figure*}
\centering
\includegraphics[width=0.8\textwidth]{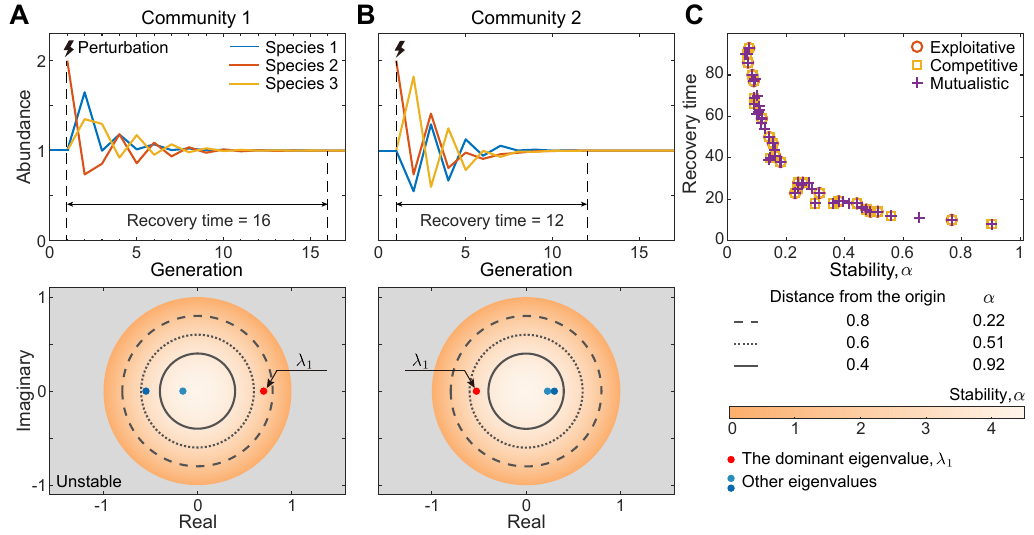}
\caption{
\textbf{Quantifying stability of discrete-time complex communities.}
\textbf{A}-\textbf{B}, Responses of three-species communities to external perturbations (top) and eigenvalue distributions of the corresponding Jacobian matrices (bottom). When the abundance of each species returns to within $0.1\%$ of the equilibrium after perturbation, we consider the community to have returned to equilibrium and calculate the recovery time accordingly. Results show that community 2 has a shorter recovery time and is thus more stable (community 1 recovers with 16 generations, community 2 recovers with 12 generations). For the stability contour plot (bottom), colors represent the intensity of stability, where darker yellow indicates lower stability intensity and grey indicates instability. Black lines with different styles are the isometric lines to the origin. Red, blue and cyan dots represent the eigenvalues of the corresponding Jacobian matrix, with the red dot pointed by an arrow denoting the dominant eigenvalue $\lambda_1$. According to our theory, the recovery time can be reflected by the distance between $\lambda_1$ and the origin: communities with shorter recovery times have a smaller distance, as shown in the bottom panels. $\alpha = - \log \left| \lambda_{1} \right|$ is introduced to quantify the community stability, and communities with higher $\alpha$ have a lower recovery time. \textbf{C}, The relationship between the recovery time and the stability $\alpha$ for exploitative, competitive, and mutualistic communities. Numerical results are obtained by perturbing three-species communities, indicating that the stability of discrete-time complex communities is quantified by $\alpha$. In A, $r_1=-0.3$, $r_2=0.9$, $r_3=0.1$, $A_{12}=-A_{21}=0.5$, $A_{13}=0.8$, $A_{23}=A_{31}=0.6$, $A_{32}=0.3$. In B, $r_1=1.3$, $r_2=-0.4$, $r_3=0.1$, $A_{12}=-A_{21}=A_{32}=0.6$, $A_{13}=A_{31}=0.3$, $A_{23}=0.8$. In A-C, $s=1$, and variables are the same as those in Fig.~\ref{overlapping_nonoverlapping}B.
}
\label{Stability_quantification}
\end{figure*}

\section{Results}
\subsection{Modeling framework}
A community composed of $S$ species with non-overlapping generations can be modeled by the famous discrete-time Lotka-Volterra dynamics \cite{may1974biological,hofbauer1987coexistence}
{\small
\begin{eqnarray}\label{diff}
&&\hspace{-0.4cm}X_{i}(k+1)=X_{i}(k)\exp\left(r_{i}-s_{i}X_{i}(k)+\sum_{j=1,j\neq i}^SA_{ij}X_{j}(k)\right)
\end{eqnarray}
}
\hspace{-0.3cm} with $i=1,2,\cdots,S,$ where $X_{i}(k)$ represents the abundance of species $i$ at generation $k$, $r_{i}$ is the intrinsic growth rate of species $i$, $s_{i}>0$ is the self-regulation strength of species $i$, and $A_{ij}$ captures the influence that species $j$ has on species $i$.

When $X^*_i>0$ ($i=1,\cdots,S$) satisfies
\begin{eqnarray*}
X^{*}_{i}=X_{i}^*\exp\left(r_{i}-s_{i}X_{i}^*+\sum_{j=1,j\neq i}^SA_{ij}X^*_{j}\right),
\end{eqnarray*}
$\mathbf{X}^*=\left[X_{1}^*,X_{2}^*,\cdots,X_{S}^*\right]^{\mathrm{T}}$ is called a feasible equilibrium, where each species has a positive abundance. Within theoretical ecology, researchers are more interested in studying community behavior around a feasible equilibrium since unfeasible equilibrium suggests species extinction occurs. The dynamical behavior around feasible equilibrium $\mathbf{X}^*$ can be approximated by linearizing equation (\ref{diff}) around $\mathbf{X}^*$ (see Supplementary Note 1)
\begin{eqnarray}\label{linear1}
&&\hspace{0cm}\Delta \mathbf{x}(k)=\mathbf{M}\Delta\mathbf{x}(k-1).
\end{eqnarray}
Here $\Delta \mathbf{x}(k)=\mathbf{X}(k)-\mathbf{X}^*$ represents the deviation from equilibrium abundance, $\mathbf{M}$ is the Jacobian matrix whose element $M_{ij}$ represents the effect that species $j$ has on species $i$ around the equilibrium.

For communities composed of species with overlapping generations (namely the continuous-time dynamics), as long as all eigenvalues $\lambda$ of the Jacobian matrix are located in the left half of the complex plane, the community is stable (Fig.~\ref{overlapping_nonoverlapping}C and Supplementary Fig.~S1). Mathematically, this requirement can be interpreted as $\max\left( \mathrm{Re}\left(\lambda\right)\right)<0$, and it can thus be seen that it is the rightmost eigenvalue that determines community stability. While for communities composed of species with non-overlapping generations (namely the discrete-time dynamics), stability requires that all eigenvalues of the Jacobian matrix lie in the unit circle centered at the origin of the complex plane (Fig.~\ref{overlapping_nonoverlapping}D and Supplementary Fig.~S1), which is equivalent to
\begin{eqnarray}\label{maxlambda}
&&\hspace{0cm}\max(|\lambda|)<1.
\end{eqnarray}
Here $|\lambda|$ is the absolute value (or magnitude) of $\lambda$, and reflects the distance between $\lambda$ and the origin of the complex plane.

It can thus be seen that, unlike communities composed of species with overlapping generations, the stability of communities with non-overlapping generations is determined by the dominant eigenvalue of the Jacobian matrix (i.e., the eigenvalue with the largest absolute value, often denoted as $\lambda_{1}$). Clearly, this difference in stability criteria can lead to different stability classifications for communities sharing the same set of community parameters but following different dynamics. As is shown in Fig.~\ref{overlapping_nonoverlapping}, a community of species with overlapping generations is stable (Fig.~\ref{overlapping_nonoverlapping}C), but its counterpart with non-overlapping generations becomes unstable (Fig.~\ref{overlapping_nonoverlapping}D). This intuitive illustration suggests that focusing solely on continuous-time dynamics can lead to misleading results when dealing with communities of species with non-overlapping generations. Consequently, a systematic and thorough exploration for the stability of communities following discrete-time dynamics is required.

\subsection{Metric for stability of discrete-time complex communities}
The unit circle stability region helps us determine whether a given community is stable, we next seek to quantify the intensity of stability. That is, when there are multiple stable communities, can we identify which community is more stable? In ecology, the intensity of stability is often assessed by recovery time (Fig.~\ref{Stability_quantification}A)--the time for a perturbation to decay to a specified fraction of its initial size \cite{pimm1984complexity,lehman2000biodiversity}. With this metric, a community with less recovery time is deemed as more stable (Fig.~\ref{Stability_quantification}A, B).

Dynamical systems theory reveals that the dominant eigenvalue $\lambda_{1}$ of the Jacobian matrix $\mathbf{M}$ not only determines community stability but also reflects the recovery time \cite{caswell2005reactivity}: the closer $\lambda_{1}$ is to the origin, the shorter the recovery time (Fig.~\ref{Stability_quantification}A, B). This is because the distance between the dominant eigenvalue and the origin of the complex plane (i.e., $\left| \lambda_{1} \right|$) dictates how fast perturbations decay over the long term: the longer the distance of $\lambda_{1}$ from the origin, the slower the decay of perturbations, and thus the longer the recovery time (see Supplementary Note 1). As a result, we next use the decay rate $\alpha=-\log \left( \left| \lambda_{1} \right| \right)$, which is developed based on the distance $\left| \lambda_{1} \right|$, to quantify the intensity of stability: the larger $\alpha$, the shorter the recovery time and, consequently, the more stable the community (Fig.~\ref{Stability_quantification}C).

Consequently, the key to determine the stability intensity is to identify the dominant eigenvalue of $\mathbf{M}$. Given that natural communities are rich in species and complex in interactions, the determination of the dominant eigenvalue can be challenging. Fortunately, this challenge can be addressed by recent advancements in random matrix theory (RMT) \cite{nguyen2015elliptic,o2014low,rogers2010universal}. For large communities, RMT points out that eigenvalues of $\mathbf{M}$ are in general distributed in an ellipse (sometimes with an outlier, see Supplementary Note 2). Therefore, four endpoints (leftmost, rightmost, uppermost, and lowermost eigenvalue) of the eigenvalue distribution are more likely to be the dominant eigenvalue (Supplementary Fig.~S3). Due to the symmetry in the eigenvalue distribution and the unit circle stability region, the uppermost and lowermost eigenvalues are equivalent. Consequently, the stability of large complex communities can be estimated as
\begin{eqnarray}\label{alpha}
\tilde{\alpha}=-\log\left\{\max\left\{\beta_1,\beta_2,\beta_3\right\}\right\}.
\end{eqnarray}
Here $\tilde{\alpha}$ is the estimated stability intensity, $\beta_1=|\mathbf{M}_{\lambda,\mathrm{leftmost}}|$, $\beta_2=|\mathbf{M}_{\lambda,\mathrm{rightmost}}|$, $\beta_3=|\mathbf{M}_{\lambda,\mathrm{uppermost}}|$. Theoretical estimations are in good agreement with numerical simulations, proving the effectiveness of our estimation theory (Supplementary Figs.~S4-S7).

\begin{figure*}
\centering
\includegraphics[width=0.7\textwidth]{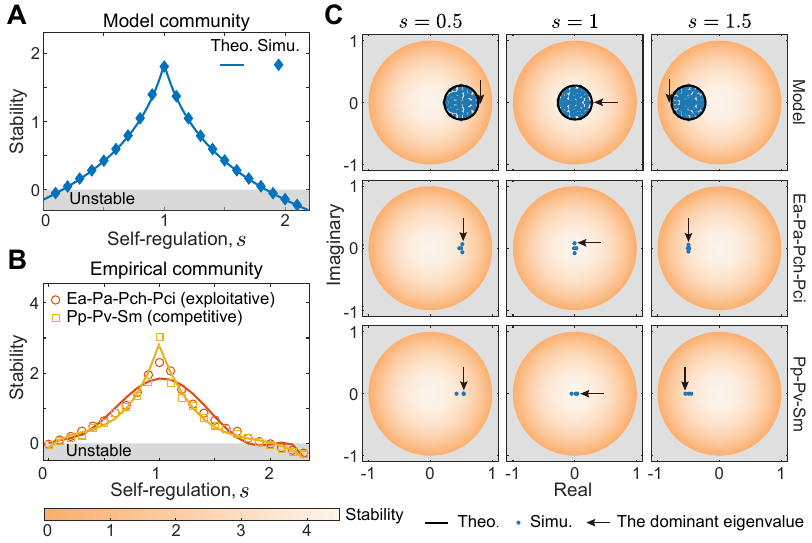}
\caption{
\textbf{Self-regulation strength modulates the ecological stability.}
\textbf{A}, The relationship between stability and self-regulation strength in communities with random interaction distribution. The solid line is the analytical results calculated by our theory (Theo.), and diamond dots are numerical simulations (Simu., each dot is an average of 50 communities with the same set of community parameters). Grey region indicates instability. \textbf{B}, Same relationship observed in empirical soil exploitative (Ea-Pa-Pch-Pci) and competitive (Pp-Pv-Sm) communities \cite{friedman2017community}. Dots are obtained by numerical simulations, and solid lines are from curve fitting. \textbf{C}, Eigenvalue distributions of model and empirical communities (Model community with random interaction distribution, top; Ea-Pa-Pch-Pci, middle; Pp-Pv-Sm, bottom) under weak ($s=0.5$), moderate ($s=1$) and strong ($s=1.5$) self-regulation. Blue dots are eigenvalues from numerical simulations (Simu.), black lines are boundaries predicted by our theory (Theo.). The blue dot pointed by the arrow represents the dominant eigenvalue (which determines stability) of the Jacobian matrix $\mathbf{M}$. In A, $\sigma=0.05$, $C=0.1$, $S=100$. In B, $\sigma=0.1$. In C, $\sigma=0.05$, $S=100$, $C=0.3$.
}
\label{selfregulation}
\end{figure*}
\subsection{Self-regulation strength modulates ecological stability}
Equipped with our theory, we next delve into the quantitative relationship between various characteristics of complex communities and ecological stability. We begin by examining the influence of the direct effects species have on themselves. Originating from mechanisms including intraspecific interference, cannibalism, time-scale separation between consumers and their resources, these self-effects decrease a species' per-capita growth when the species experiences an abundance increase, and are thus known as `self-regulation' \cite{may2007theoretical,allesina2012nature,barabas2017nee}. Previous studies on communities composed of species with overlapping generations revealed that self-regulation plays a pivotal role in regulating stability: communities can uniformly benefit from the increase of self-regulation strength, and ecological stability can be maintained as long as species exhibit substantially strong self-regulation \cite{barabas2017nee}.

Surprisingly, we find that these established rules do not apply for communities composed of species with non-overlapping generations. Our analyses indicate that, in such communities, the relationship between self-regulation strength and stability is non-monotonic: increasing self-regulation strength initially enhances community stability, then inhibits it, and ultimately leads to instability (Fig.~\ref{selfregulation}A, Supplementary Figs.~S5-S7). Furthermore, we select two empirical microbial communities isolated from the soil \cite{friedman2017community}. Community 1 is an exploitative community, and is composed of \emph{E. aerogenes} (Ea), \emph{P. aurantiaca} (Pa), \emph{P. chlororaphis} (Pch), and \emph{P. citronellolis} (Pci). Community 2 is a competitive community, and is composed of \emph{P. putida} (Pp), \emph{P. veronii} (Pv), and \emph{S. marcescens} (Sm). We confirm that the non-monotonic relationship also exists in these empirical communities (Fig.~\ref{selfregulation}B, see Supplementary Note 3).

The above results can be intuitively understood by examining the eigenvalue distribution under different self-regulation strengths and the unit circle stability region (Fig.~\ref{selfregulation}C). Taking communities with random interaction distribution (such communities encompass exploitative, competitive, and mutualistic interactions, and these three types of interactions are `well-mixed') as an example. We show that when self-regulation strength is relatively low, the rightmost eigenvalue is the dominant eigenvalue $\lambda_1$ (top row of Fig.~\ref{selfregulation}C). As self-regulation strength increases, the rightmost eigenvalue shifts toward the left, leading to a reduction in its absolute value (i.e., it becomes closer to the origin of the complex plane), thereby enhancing community stability. Conversely, when self-regulation strength is relatively high, the leftmost eigenvalue becomes the dominant eigenvalue (right panel). Further increases in self-regulation strength result in the leftmost eigenvalue shifting further leftward, leading to an increase in its absolute value (i.e., it moves away from the origin of the complex plane), thus inhibiting community stability. Similar patterns are also found in selected empirical soil microbial communities with different compositions of interaction types (Middle and bottom rows of Fig.~\ref{selfregulation}C).

Stemming from this non-monotonic relationship, another important and surprising finding is that maintaining ecological stability imposes a stricter requirement on self-regulation strength: it must reside within a suitable range. Too strong or too weak self-regulation can both lead to instability for communities composed of species with non-overlapping generations (Fig.~\ref{selfregulation}A, B, Supplementary Figs.~S5-S7).

\begin{figure*}
\centering
\includegraphics[width=\textwidth]{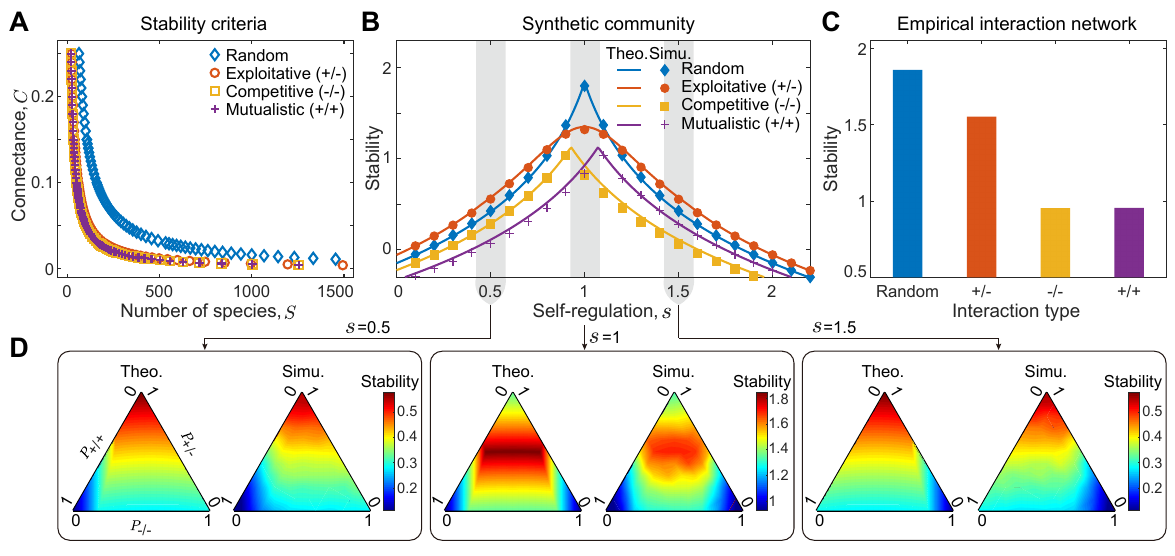}
\caption{
\textbf{Stability of complex communities with typical interaction types.} \textbf{A}, Stability criteria for communities with different types of interactions. Different markers represent the critical $S-C$ curves for different communities, and combinations of $S$ and $C$ below each curve result in stable communities with a probability close to 1. Different types of communities present a hierarchy, from random communities (most likely to be stable) to mutualistic/competitive communities (most likely to be unstable). \textbf{B}, The relationship between stability and self-regulation strength for different communities. Solid lines are obtained from our theory (Theo.) and dots are from numerical simulations (Simu.). \textbf{C}, Stability analyses of typical communities with empirical interaction network. We extract the interaction network from the mouse gut microbiome \cite{stein2013ecological}. For each type of community (random, exploitative, competitive, and mutualistic), we generate 100 samples respectively and calculate the average stability. \textbf{D}, Stability of communities with diverse interaction types: exploitation, mutualism and competition with proportion $P_{+/-}$, $P_{+/+}$, and $P_{-/-}$, respectively. Left part of each panel gives results from our theory (Theo.) and right shows numerical simulations (Simu.). Each data point in the right part is an average of 50 randomly generated communities with the same set of parameters. In A, $s=1$, and $\sigma=0.05$, $C=0.1$, $S=100$ in B and D.
}
\label{interaction_types}
\end{figure*}
\subsection{Communities can benefit from the diversity of interaction types}
For complex communities, pairs of species can interact in a range of well-defined ways, such as exploitative, competitive, and mutualistic interactions. Disentangling influences of different types of interactions on stability is a fundamental focus in ecology \cite{allesina2012nature,mougi2012diversity,tang2014correlation,barabas2017nee,coyte2015ecology}. Previous studies based on continuous-time dynamics indicate that the stability of these four typical communities--random, exploitative, competitive, and mutualistic--exhibits a strict hierarchical order: exploitative communities are the most stable, followed by random communities, then competitive communities, and mutualistic communities are the least stable \cite{allesina2012nature}. This suggests that exploitative interactions are stabilizing, while competitive and mutualistic interactions are destabilizing. Moreover, conventional work suggests that this ranking is robust for other community traits, such as self-regulation \cite{allesina2012nature}, short delays \cite{yang2023time}, abundance \cite{yang2023time}, suggesting that ecological stability can consistently benefit from exploitative interactions. We then wonder if these results still hold for communities of species with non-overlapping generations.

Unexpectedly, we find that the stability of communities composed of species with non-overlapping generations benefits from the diversity of interaction types (Fig.~\ref{interaction_types}A), in contrast to the continuous-time counterparts, which consistently benefit from exploitative interactions (Fig. S9). As is shown in Fig.~\ref{interaction_types}A, random communities--where different types of interactions are `well-mixed'--exhibit the highest stability performance among these four typical communities. Exploitative communities now rank second, while competitive and mutualistic communities possess the worst stability performance.

By checking the eigenvalue distribution of each type of communities, this stability pattern can be intuitively understood in the context of our theory. In case of all species self-regulate with unit strength, the eigenvalue distributions of all communities are centered at the origin. For random communities, the eigenvalues are distributed in a circle with radius $r$, suggesting that the distance between the origin and the dominant eigenvalue is $r$ (Supplementary Note 4 and Fig.~S10). For exploitative communities, the eigenvalues are distributed in a vertically stretched ellipse with a semi-horizontal axis $r_{\text{ex}}<r$ and a semi-vertical axis $r_{\text{ey}}>r$ (Supplementary Note 4 and Fig.~S10). Therefore, the distance between the origin and the dominant eigenvalue is $r_{\text{ey}}$, which is greater than that of random communities, resulting in the decrease of stability. For competitive/mutualistic communities, the eigenvalues can be divided into two parts: a horizontally stretched ellipse and an outlier on the horizontal axis outside this ellipse (Supplementary Note 4 and Fig.~S10). This outlier becomes the dominant eigenvalue for competitive/mutualistic communities, and its distance from the origin is greater than that of exploitative communities. As a result, these two types of communities exhibit the lowest stability performance.

With the exception of recognizing that the ecological stability with non-overlapping generations can be promoted by introducing the diversity of interaction types, we further explore under what circumstances this promotion can occur. We find that this is modulated by self-regulation strength. With a low level of self-regulation, the stability order of these typical communities keeps the traditional ranking from continuous-time dynamics (Fig.~\ref{interaction_types}B). With a high level of self-regulation, exploitative communities reclaim their position as the most stable case, followed by random communities. Mutualistic communities occupy the third spot, while competitive communities lag behind (Fig.~\ref{interaction_types}B). With a moderate level of self-regulation, random communities exhibit the highest stability performance (Fig.~\ref{interaction_types}B), as we have shown in Fig.~\ref{interaction_types}A.

\begin{figure*}
\centering
\includegraphics[width=0.9\textwidth]{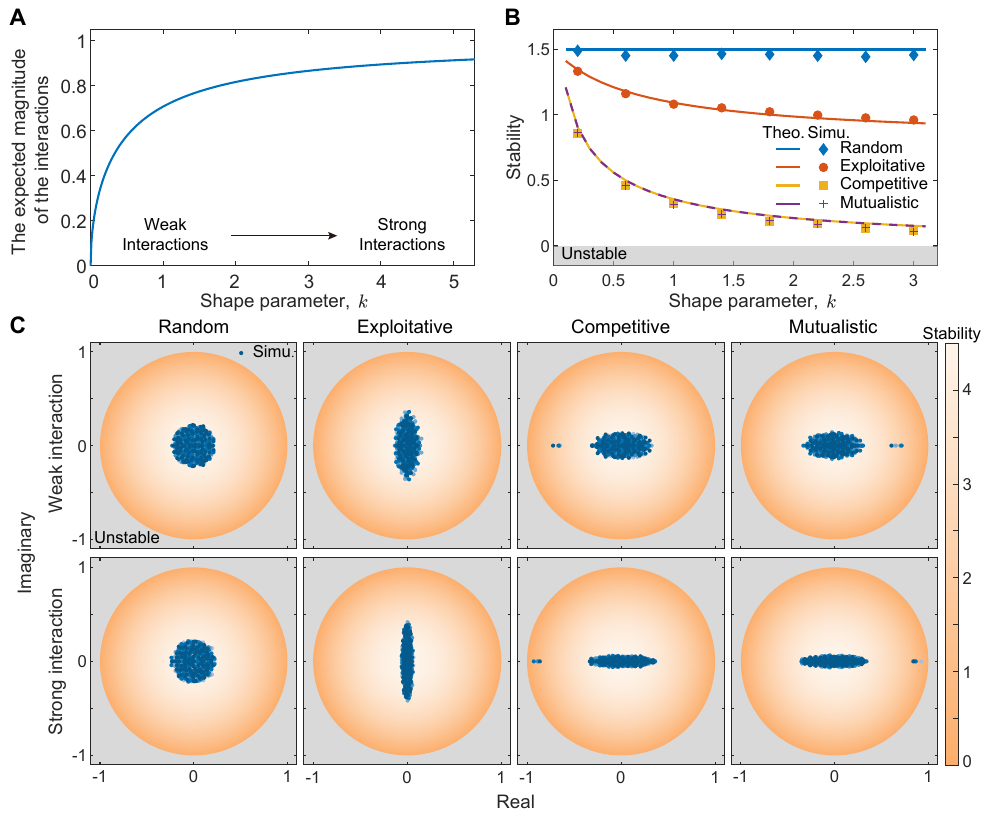}
\caption{
\textbf{Weak interactions tend to stabilize complex communities.}
\textbf{A}, Shape parameter $k$ controls the prevalence of weak interactions. The expected magnitude of interactions reflects the prevalence of weak interactions, where rare weak interactions result in high magnitude, while prevalent weak interactions result in low magnitude. The increase of $k$ leads to the increase of the expected magnitude of interactions, and thus the increase of weak interactions prevalence. \textbf{B}, The relationship between stability and $k$ for typical communities (random, exploitative, competitive, and mutualistic). Solid lines are theoretical calculations (Theo.), and dots are results from numerical simulations (Simu.). \textbf{C}, Eigenvalue distributions of Jacobian matrices for different types of communities under weak ($k=0.7$, top) and strong ($k=3$, bottom) interactions. Blue dots are eigenvalues from numerical simulations. Other settings are the same as those in Fig.~\ref{selfregulation}C except that $C=0.2$.
}
\label{weak_interactions}
\end{figure*}

Apart from corroborating our findings through theoretical analyses and numerical simulations, we test our key predictions with recently published data on mouse gut microbiome \cite{stein2013ecological}. We extract interaction networks from the data to generate corresponding random, exploitative, competitive, and mutualistic communities, respectively. We confirm that this surprising stability pattern emerges not only in synthetic communities but also in communities constructed from empirical data (Fig.~\ref{interaction_types}C). This suggests that the stability promotion driven by the diversity of interaction types occurs when species have a moderate level of self-regulation. Further extensions to communities where different types of interactions are mixed with arbitrary proportions further support this finding from theory and numerical simulations (Fig.~\ref{interaction_types}D).

\subsection{Weak interactions tend to stabilize complex communities under discrete-time dynamics}
Other than interaction types, the distribution of interaction strength plays a critical role in characterizing species interactions \cite{may2001stability,may1972nature,may2007theoretical,allesina2012nature,allesina2015stability,mccann1998weak,ratzke2020strength,hu2022emergent}. In communities of species with overlapping generations, it is found that the influence of interaction strength distribution on stability depends on specific interaction types: the prevalence of weak interactions can destabilize exploitative communities, can stabilize competitive and mutualistic communities, and has no influence on random communities \cite{allesina2012nature}. To adjust the prevalence of weak interactions in complex communities, an effective and efficient technical way is to sample interaction strengths from a Gamma and a reflected Gamma distribution \cite{allesina2012nature}, which allows us to tune the prevalence of weak interactions in a given community by a single parameter (Fig.~\ref{weak_interactions}A, see Supplementary Note 5).

Our analyses reveal that, weak interactions can in general promote community stability except for random communities, which appears to be insensitive to the prevalence of weak interactions (Fig.~\ref{weak_interactions}B). This is because for exploitative, competitive, and mutualistic communities, the prevalence of weak interactions affects the eigenvalue distribution of the Jacobian matrix, and further leads to the decrease of the magnitude of the dominant eigenvalue (i.e., stability is enhanced, Fig.~\ref{weak_interactions}C). While for random communities, the prevalence of weak interactions has no influence on the eigenvalue distribution of the Jacobian matrix, and thus has no influence on stability (Fig.~\ref{weak_interactions}C). Note that these results are consistent with recent experimental findings showing that weak interactions can stabilize microbial communities \cite{ratzke2020strength,hu2022emergent}.

\subsection{Structured food webs composed of species with non-overlapping generations}
So far, our theoretical analyses have primarily focused on communities with unstructured interaction networks (i.e., species interact randomly), implying that the occurrence of interactions between any two species is random. However, supported by empirical data, interaction networks of real-world ecological communities can be structured. For example, interaction networks of empirical food webs often exhibit hierarchical structure (i.e., trophic levels), intervality, and broad degree distribution \cite{chen2001transient,stouffer2006robust,dunne2002food,williams2000simple,allesina2008general,cohen2012community} (here degree of a species captures the number of its interacting partners). We are then interested in how these structures influence stability when species have non-overlapping generations.

\begin{figure*}
\centering
\includegraphics[width=0.8\textwidth]{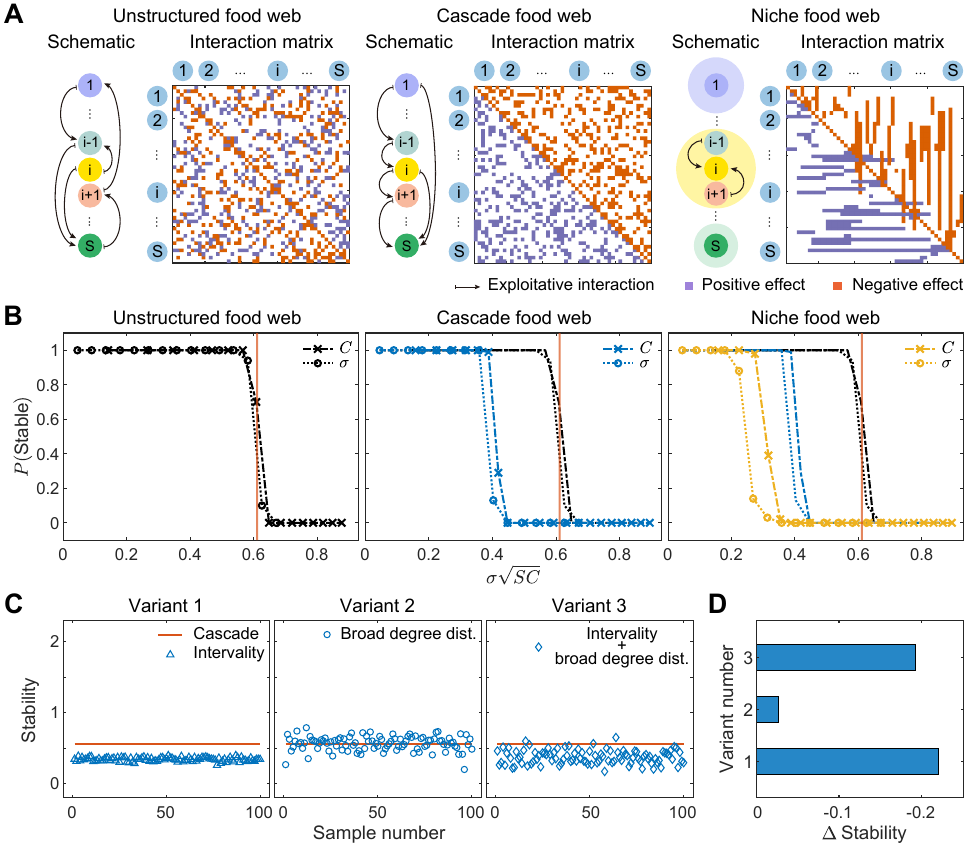}
\caption{
\textbf{Stability of structured food webs with non-overlapping generations.}
\textbf{A}, Diagrams of interaction network structure and interaction matrices. In the unstructured food web, two species interact with each other randomly, and the role of predator or prey is assigned with equal probability. In the cascade food web, ranks (numbers) of species are introduced, and species with higher ranks prey on species with lower ranks with a fixed probability. And the curved arrow indicates exploitative interaction. In the niche food web, species prey on all species within its predation range. For example, the yellow species $i$ preys on all species located within the yellow--shaded circle. Purple and red blocks in interaction matrices represent positive and negative interactions, respectively. \textbf{B}, Transition from stability to instability for food webs. We systematically vary the connectance $C$ (cross) or the standard deviation $\sigma$ (circle) to obtain the critical value of $\sigma \sqrt{SC}$ for instability. The vertical red line is the critical value calculated by our theory for unstructured food web. Black, blue, and yellow lines and markers (crosses and circles) are numerical results, respectively. Each marker is calculated over 100 randomly generated communities with the same set of parameters. \textbf{C}, Stability of the cascade food web with different variants. Variant 1 is interval cascade food web; variant 2 is cascade food web with a broad degree distribution; variant 3 is a combination of 1 and 2. Red lines indicate the stability of the cascade food web without these variants and blue markers (triangle, circle, and diamond) represent the stability with different variants. For each type of variant, we randomly generated 100 communities with the same parameters. \textbf{D}, Average stability change for different variants shown in panel C, and here $\Delta \mathrm{Stability} = \mathrm{Stability}_{\mathrm{Variant}} - \mathrm{Stability}_{\mathrm{Cascade}}$. In C-D, $S=200$, $C=0.1$, $s=1$, $\sigma=0.05$.
}
\label{structured}
\end{figure*}

To address this problem, here we construct model exploitative communities following two widely used food web models -- cascade model \cite{allesina2012nature,barabas2017nee,allesina2015predicting,chen2001transient,allesina2008general,cohen2012community} and niche model \cite{allesina2012nature,barabas2017nee,allesina2015predicting,mcnaughton1978stability,yodzis1981stability} (see Supplementary Note 6). In the cascade model, species are organized into a strict hierarchy, where species occupying higher ranks predate species positioned at lower ranks with a fixed probability (Fig.~\ref{structured}A). Therefore, cascade model introduces trophic levels compared with unstructured food web. In the niche model, each species is assigned a specific predation range, and accordingly preys upon all species that fall within this range (Fig.~\ref{structured}A). Compared with cascade model, niche model further introduces intervality and broad degree distribution \cite{yang2023reactivity}.

Compared with unstructured food webs, we find that both cascade and niche food webs exhibit reduced stability, suggesting that incorporating structural features of empirical food webs tends to exert a destabilizing effect (Fig.~\ref{structured}B and Supplementary Fig.~S14). The observed decline in stability from unstructured to cascade food webs indicates that the introduction of trophic levels is a destabilizing factor. The additional decrease in stability seen when transitioning from cascade to niche food webs highlights the destabilizing impact of introducing intervality (each predator consumes preys that are adjacent in the hierarchy) alongside a broad degree distribution within food web architectures. By constructing variants of cascade model (interval cascade model, cascade model with broad degree distribution, and interval cascade model with broad degree distribution, see Supplementary Note 6), we identify that intervality is the key driver for the stability decrease from cascade model to niche model (Fig.~\ref{structured}C,D and Supplementary Fig.~S15). This is because intervality causes the dominant eigenvalues to be further away from the origin, thus reducing stability (Supplementary Fig.~S16).

\section{Discussion}

Since May's pioneering work, ecologists have dedicated tremendous effort to understanding the mechanisms driving stability or instability in complex ecological communities. However, most existing theories focus on continuous-time dynamics, rendering them ineffective in explaining the behavior of communities composed of species with non-overlapping generations. To address this gap, here we develop a discrete-time dynamical framework to comprehensively understand the drivers of ecological stability in these communities. With this framework, we reveal that several established rules for ecological stability need to be revised. It is worth noting that although the main focus of our work is stability, this framework can also be applied to study other critical properties of complex communities, such as coexistence \cite{saavedra2017structural}, and reactivity \cite{caswell2005reactivity,neubert1997alternatives,tang2014reactivity,yang2023reactivity}, to name a few.

In stark contrast to classic research on species communities with overlapping generations, our study reveals that self-regulation plays a more versatile role in shaping the stability of complex communities with non-overlapping generations. Specifically, self-regulation acts as a double-edged sword: increasing its strength is initially stabilizing but eventually becomes destabilizing. This dual effect cautions ecosystem managers against assuming that stronger self-regulation is always beneficial. Moreover, self-regulation modulates the influence of different interaction types on stability, and communities benefit most from interaction-type diversity when species exhibit a moderate level of self-regulation. This finding highlights that the advantage of diverse interaction types is a key feature of natural communities in maintaining ecological stability.

Apart from revealing the beneficial influence of diverse interaction types, the interplay between self-regulation and interspecies interactions also provides insights into whether populations tend to prefer competition or cooperation (mutualism)--a long-standing and debated topic in microbiology. Some microbiologists argue that bacterial species rarely cooperate, as cooperation may undermine community stability \cite{coyte2015ecology,oliveira2014evolutionary}. Others, however, contend that cooperation enhances stability and is therefore favored \cite{ona2022cooperation,d2016experimental}. Given that many bacterial species reproduce through binary fission--a process in which a single cell divides into two genetically identical daughter cells and inherently follows a discrete-time structure--our theory may help reconcile these seemingly contradictory perspectives: when communities exhibit low or high levels of self-regulation, competitive or mutualistic interactions, respectively, lead to higher level of stability.

Building on our current work and classic studies, a natural next step is to investigate how to model and analyze the dynamical behavior of complex communities that contain a mixture of species with overlapping and non-overlapping generations. This is a valuable direction, as such communities are more representative of natural systems than those composed solely of species with overlapping or non-overlapping generations. However, the fact that modeling these communities requires combining continuous-time and discrete-time frameworks makes the exploration of this topic particularly challenging.

In general, our study highlights that established rules for ecological stability may falter when applied to communities composed of species with non-overlapping generations, whose behavior is governed by discrete-time dynamics. Consequently, without a thorough understanding of the stability mechanisms in such communities, ecosystem managers may implement ill-advised measures in response to ecological crises--potentially leading to catastrophic consequences for the environment, climate, and the routine functioning of human societies. Although our analyses focus on communities with non-overlapping generations, the results can be broadly interpreted and extended to a wide range of systems described by difference equations. Moreover, since real-world data are often collected in discrete time, our theory can even be applied to communities with overlapping generations. Thus, our work offers a potential bridge between theory and empirical research, contributing to the understanding and management of ecological communities.

\bibliography{references}

\end{document}